\documentclass[submission,copyright,creativecommons]{eptcs}
\providecommand{\event}{GCM 2020} % Name of the event you are submitting to
\usepackage{breakurl}             % Not needed if you use pdflatex only.
\usepackage{macros}
\usepackage{graphicx}
\usepackage{amsmath,amssymb}
\usepackage{multicol}
\usepackage{tikz}
\usepackage{tikz-cd}
\usepackage{verbatim}
\usepackage{wrapfig}
\usepackage{subcaption}
\usepackage{stmaryrd}
\usepackage{trimclip}
\usepackage{mathtools}
\usepackage{relsize}
\usepackage{wrapfig}

\usetikzlibrary{shapes, snakes, backgrounds, calc}
\usetikzlibrary{decorations.pathmorphing}
\usepackage{transparent}

\def\from{\leftarrow}

\def\to{\rightarrow}
\def\into{\rightarrowtail}

\newenvironment{diagram}[1][row sep=normal, column sep=normal]{
	\begin{equation}
	\begin{tikzcd}[#1, ampersand replacement=\&]
}
{ 
	\end{tikzcd} 
	\end{equation}
}

\title{Reversibility and Composition of Rewriting in Hierarchies}
\author{Russ Harmer
\institute{Univ Lyon, EnsL, UCBL, CNRS, LIP}
\institute{F-69342 LYON Cedex 07, France}
\email{russell.harmer@ens-lyon.fr}
\and
Eugenia Oshurko
\institute{Univ Lyon, EnsL, UCBL, CNRS, LIP}
\institute{F-69342 LYON Cedex 07, France}
\email{\quad ievgeniia.oshurko@ens-lyon.fr}
}
\def\titlerunning{Reversibility and composition of rewriting in hierarchies}
\def\authorrunning{R. Harmer and E. Oshurko}
\begin{document}
\maketitle

\begin{abstract}

In this paper, we study how graph transformations based on sesqui-pushout rewriting can be reversed and how the composition of rewrites can be constructed. We illustrate how such reversibility and composition can be used to design an audit trail system for individual graphs and graph hierarchies. This provides us with a compact way to maintain the history of updates of an object, including its multiple versions. The main application of the designed framework is an audit trail of updates to knowledge represented by hierarchies of graphs. Therefore, we introduce the notion of rule hierarchy that represents a transformation of the entire hierarchy, study how rule hierarchies can be applied to hierarchies and analyse the conditions under which this application is reversible. We then present a theory for constructing the composition of consecutive hierarchy rewrites. The prototype audit trail system for transformations in hierarchies of simple graphs with attributes is implemented as part of the ReGraph Python library.

\end{abstract}

\section{Introduction}

The main goal of this work is to design a system for maintaining an \emph{audit trail} for the knowledge representation (KR) framework based on hierarchies of graphs presented in \cite{harmer2020knowledge}. We would like to use this system to record the history of predominantly small localized updates of a large knowledge corpus represented with graphs and graph hierarchies, where storing the corpus at each point in the history is not feasible. The aforementioned requirement for the transformations to be reversible in the context of our practical application does not restrict the set of operations that can be performed for updating a corpus, but rather introduces a reasonably small overhead on storing additional information that allows us to revert the applied transformations. Furthermore, we design means for maintaining multiple versions of the corpus, crucial when accommodating different versions of knowledge that, for example, correspond to different view-points of knowledge curators or some intrinsic knowledge conflicts (e.g. contradicting experimental results, alternative hypotheses). The main use cases of audit trails for KRs based on hierarchies of graphs include version control for updates in schema-aware graph databases presented in \cite{bonifati2019schema} and updates to the knowledge corpora provided by the bio-curation framework 
KAMI \cite{harmer2019bio}. 

An audit trail system with the desired capabilities heavily relies on: (1) the existence of an \emph{efficient semantic representation of object transformations} (or deltas), which frees us from the necessity to store the state of the object at each point of its transformation history; (2) the \emph{reversibility of transformations}, which guarantees that any sequence of transformations can be `undone'; and (3) the existence of sound means for \emph{composing a pair of successive transformations}, which allows us to efficiently store and switch between different versions of the same object. In this work we formulate the three above-mentioned ingredients with respect to both individual objects (Section \ref{s:preliminaries}), for instance graphs, and hierarchies of objects (Section \ref{s:rulehierarchy}) and discuss how these ingredients can be used to construct an efficient audit trail (Section \ref{s:audit}).

We present a mathematical framework for building a traceable history of transformations based on the sesqui-pushout (SqPO) approach \cite{corradini2006sesqui} operating on individual objects and hierarchies of objects from appropriately structured categories. A transparent audit trail provides insight into the history of object transformations and allows us to revert to an arbitrary point in this history. The latter feature is extremely useful, when, for example, trying to fix an erroneous transformation. Moreover, such an audit trail provides means for efficient accommodation of multiple versions of the same object arising, for example, as the result of inconsistent transformations.

\subsection*{Related work}

SqPO rewriting is a generalization of the double (DPO) and single (SPO) pushout approaches to graph transformation that allows for the cloning of graph elements in addition to the familiar operations of adding, deleting and merging of graph elements.
While the reversibility of SqPO rewriting has been previously studied \cite{danos2014reversible}, the composition of consecutive (and not necessarily sequentially independent) SqPO rewrites has been less fully developed. The construction of such compositions has been given for the DPO approach (see D-concurrent productions in \cite{ehrig1991parallelism}) and for special cases of the SqPO approach (where rules are linear \cite{behr2019sesqui} and where the right-hand side of the first applied rule is exactly the left-hand side of the second rule \cite{lowe2015polymorphic}). In this work, we present such a construction for two consecutive applications of general SqPO rules, where the first is required to be reversible. This was inspired by \cite{ehrig1991parallelism} and adapted to SqPO rewriting. For the composition to be well-defined, we require the underlying category to have structure allowing for well-behaved generalized unions of objects (more formally, we require for our category to be adhesive \cite{lack2005adhesive}). The notion of rule hierarchy and the question of the reversibility and composition of hierarchy rewrites represent a novel direction that generalizes SqPO rewriting of individual objects to hierarchies of objects \cite{harmer2020knowledge}.

The main application of interest to us, a transformation audit trail, is closely related to the version control systems (VCSs) used in software development. While such systems typically provide control over different versions of software source code (represented by text files), our audit trail provides such control for different versions of a graph or a graph hierarchy. Similarly to VCSs, the transformation audit trail avoids maintaining the state of an object at the time of every transformation by keeping only its current state together with a compact representation of a history of transformations. Moreover, by using a mechanism similar to \emph{delta compression} in VCSs, such audit trails allow for the maintenance of multiple versions of the same object and for switching between these versions.

The versioning of diagrams expressing object-oriented data modes in the domain of model-driven engineering is closely related to audit trails for graph transformations because such model diagrams can be seen as graphs with attributes. However, work in this domain \cite{schneider2004coobra, barmpis2015towards} typically considers the operations of addition and deletion of graph elements (including attributes), to express changes to the data model, but largely omits those of cloning and merging, as these operations are hard to interpret in the context of data modelling, whereas these operations are highly interesting in the context of KR.

There exists an extensive body of work on versioning for database systems \cite{salzberg1999comparison}, including graph databases\footnote{Versioning for the Neo4j database \url{https://github.com/h-omer/neo4j-versioner-core}} \cite{castelltort2013representing, khurana2013efficient}. Such systems are usually divided into snapshot- and delta-based classes (also called copy and log), where the first store snapshots of the entire database at different points of the revision history, while the second preserve the transformation log. In this context, our system falls into a rather hybrid category: while the audit trail stores a representations of transformations (logs), these logs themselves consists of subgraphs, i.e.\, graph patterns that are affected as the result of transformation. 

A different classification of versioning systems for graph databases divides them into those that provide the revision history from the point of view of individual nodes and edges (or relationships), and those that use the graph point of view \cite{castelltort2013representing}. While existing tools, such as \texttt{neo4j-versioner-core} or the one presented in \cite{castelltort2013representing}, mostly tackle versioning from the point of view of individual elements, our framework allows us to version the entire graph and even the hierarchy of graphs. Moreover, as is the case with model versioning, existing database versioning techniques do not allow to record the operations of cloning and merging of graph elements.

\section{Preliminaries}\label{s:preliminaries}

In this section we briefly present some useful notions that serve as preliminaries for the rest of the paper and allow us to construct the desired audit trail for SqPO rewriting of individual objects. We introduce SqPO rewriting, its reversible version and add the audit trail ingredient by presenting in Section \ref{ss:composition} how the composition of two consecutive SqPO rewrites can be constructed when the first rewrite is reversible. Finally, we conclude this section by presenting the principal KR model of interest, a hierarchy of objects. The preliminaries, as well as the other technical parts of the paper, are presented using the rather formal notions of pullback, pushout and pullback complement originating from category theory, which can be thought of as generalized intersection, union and difference of objects (e.g. graphs) respectively.

\subsection{SqPO Rewriting}

% \eugenia{the abstract formulation of SqPO rewriting requires the categorical notion of a final pullback complement (PBC). it is not classical as pushout or pullback, we, therfore, give the definiton...}

\begin{wrapfigure}{r}{0.33 \textwidth}
\vspace{-15pt}
\hspace{-10pt}
\begin{equation}\label{d:sqprewriting}
\begin{tikzcd}[ampersand replacement=\&]
	L \arrow[dr, phantom, "(a)"] \arrow[d, tail, "m"] \& P  \arrow[l, "r^-"'] \arrow[d, tail, "m^-"] \arrow[r, "r^+"] \arrow[dr, phantom, "\ (b)"]  \& R \arrow[d, tail, "m^+"] \\
	G \& G^- \arrow[l, "g^-"] \arrow[r, "g^+"'] \& G^+ \\
\end{tikzcd}\hspace{-10pt}
\end{equation}
\vspace{-30pt}
\end{wrapfigure}
SqPO rewriting is an approach to abstract deterministic rewriting in any category with POs and (final) pullback complements (PBCs) over monos \cite{corradini2006sesqui}. In typical concrete settings, it enables us to apply operations of addition, deletion, cloning and merging of elements where, by element, we mean any concrete constituent of an object in a category of interest (such as nodes and edges in categories of graphs). The rewrite of an object $G$ is defined by a \emph{rule} $r: L \lfrom{r^-} P \lto{r^+} R$ and its \emph{instance} given by a mono $m: L \into G$. The application of $r$ is performed in two phases as in Diagram \ref{d:sqprewriting}: (a) an object $G^{-}$ is constructed as the PBC of $r^-$ and $m$ and (b) the final result of rewriting $G^+$ is constructed as the PO of $m^-$ and $r^+$. An arbitrary rewrite of an object in a category of interest can be decoupled into two phases: the \emph{restrictive} rewrite (Diagram \ref{d:sqprewriting}a) performing deletion and cloning of elements and the \emph{expansive} rewrite (Diagram \ref{d:sqprewriting}b) performing merging and addition of elements.

Transformations of individual objects through SqPO rewriting can be efficiently represented with corresponding rewriting rules and their instances. However, such rewriting may introduce \emph{side-effects}, i.e. graph transformations not explicitly specified by the underlying rules and instances. The nature of these side-effects depends on the category in which we are working. For example, in both simple and non-simple graphs, edges not matched by the left-hand side of the rule can be removed as a side-effect of a node removal. Due to such side-effects, having applied a rewriting rule to an object, we can no longer restore this object by simply looking at the applied rule and its instance. The reversible variant of SqPO rewriting that does not introduce side-effects was presented in \cite{danos2014reversible}. It corresponds to the scenario where the SqPO rewriting diagram can be read both forwards and backwards. More formally:

\begin{definition1}\label{def:reversibleSqp} An SqPO rewriting as in Diagram  1 is \emph{reversible}, if the square (a) is also a PO and the square (b) is also a PBC, i.e. $P \linto{m^-}$ $G^- \lto{g^+} G^+$ is the final PBC of $r^+$ and $m^+$. We call $r^{-1}: R \lfrom{r^+} P \lto{r^-} L$ the \emph{reverse} of $r$.
\end{definition1}

\begin{remark}
When restricting SqPO rewriting to its reversible variant, we lose the capability to perform deletions in an unknown context \cite{corradini2006sesqui}. However, this does not reduce our rewriting to the DPO approach \cite{corradini1997algebraic} due to the operation of cloning, allowed by the PBC constituting the first phase of SqPO but not realizable by DPO (or SPO). In the context of KR, the operation of cloning represents an important update capability which allows us to perform concept refinement, as discussed in \cite{bonifati2019schema} and \cite{harmer2020knowledge}.
\end{remark}

\subsection{Composition of SqPO Rewriting}\label{ss:composition}

%In this subsection we would like to study how consecutive applications of SqPO rules can be composed into a single rewrite. 
Let $r_1: L_1 \from P_1 \to R_1$ be a rewriting rule applied to an object $G_1$ through an instance $m_1: L_1 \into G_1$ and let $G_2$ be the result of application of this rule (corresponding to Diagram \ref{composition:d1}). Let $r_2: L_2 \from P_2 \to R_2$ be a rule applied to the resulting object $G_2$ through an instance $m_2: L_2 \into G_2$ (as in Diagram \ref{composition:d2}).

\vspace{-5pt}

\hspace{-20pt}
\begin{minipage}[t]{0.5\textwidth}
\begin{diagram}\label{composition:d1}
		L_1 \arrow[d, tail, "m_1"] \& P_1 \arrow[d, tail, "m_1^-"] \arrow[l, "r^-_1"'] \arrow[r, "r^+_1"] \&  R_1 \arrow[d, tail,  "m_1^+"'] \\
		G_1 \& G^-_1 \arrow[l, "g_1^-"] \arrow[r, "g_1^+"'] \& G_2 \\
\end{diagram}
\end{minipage}
\begin{minipage}[t]{0.5\textwidth}
\begin{diagram}\label{composition:d2}
	 L_2 \arrow[d, tail, "m_2"] \& P_2 \arrow[d,tail,  "m_2^-"] \arrow[l,  "r_2^-"'] \arrow[r, "r_2^+"] \& R_2 \arrow[d, tail,  "m_2^+"] \\
		G_2 \&  G^-_2 \arrow[l, , "g_2^-"] \arrow[r, , "g_2^+"'] \& G_3 \\
\end{diagram}
\end{minipage}

\vspace{-15pt}

\begin{wrapfigure}{r}{0.33\textwidth}
\vspace{-15pt}
\begin{equation}\label{d:synthesis}
\hspace{0pt}
\begin{tikzcd}[ampersand replacement=\&]
    L \arrow[d, tail, "m"] \& P \arrow[d, tail, "m^-"] \arrow[l, "r^+"'] \arrow[r, "r^-"] \&  R \arrow[d, tail, "m^+"] \\
	G_1 \& G_1^{\ominus} \arrow[l, "g^-"] \arrow[r, "g^+"'] \& G_3 \\
\end{tikzcd}\hspace{-10pt}
\end{equation}
\vspace{-35pt}
\end{wrapfigure}
Given these two consecutive rule applications, we would like to find a rule $L \lfrom{r^-} P \lto{r^+} R$ and an instance $m: L \into G_1$ that, when applied to $G_1$, directly produces the object $G_3$, i.e. such that Diagram \ref{d:synthesis} is an SqPO diagram. Apart from being well-structured for SqPO rewriting, the construction of such a rule will require the category in which we are working to be \emph{adhesive} \cite{lack2005adhesive}. 

Let us first proceed by constructing the pullback (PB) $R_1 \lfromin{x} D \linto{y} L_2$ from $m_1^+$ and $m_2$.
% as in Diagram \ref{composition:d11}. 
Note that, because PBs preserve monos, arrows $x$ and $y$ are monos. We will call the span given by this PB the \emph{overlap} of $R_1$ and $L_2$ given their matching inside $G_2$, and we will denote it with $o$. The PB of $x$ and $r_1^+$ indicates whether the two rule applications are \emph{sequentially independent}: if this PB is isomorphic to $D$ then the overlap is included in the preserved region $P_1$ of $r_1$ and the application of $r_2$ did not depend on the prior application of $r_1$; otherwise, the application of $r_1$ creates something that is tested by $r_2$ so that its application is dependent on the prior application of $r_1$ (see \cite{danos2014reversible,harmer2017hdr} for more formal details).
\begin{wrapfigure}{r}{0.33 \textwidth}
\vspace{-5pt}
\begin{diagram}[column sep=small, row sep=small]\label{composition:d12}
		      \& D \arrow[dl, tail, "x"'] \arrow[dr, tail, "y"] \\
	 R_1  \arrow[dr, tail, "r^H_1"'] \arrow[ddr, bend right=35, tail, "m_1^+"'] \& \& L_2 \arrow[ddl, bend left=35, tail, "m_2"] \arrow[dl, tail, "l^H_2"] \\
		    \& H \arrow[d, tail, dashed, "\ m^H"] \\
		    \& G_2 \\		      
\end{diagram}
%\vspace{-25pt}
\end{wrapfigure}
The PO $R_1 \linto{l^H_1} H \lfromin{l^H_2} L_2$ from $x$ and $y$ as in Diagram \ref{composition:d12} constructs the object $H$ that can be seen as the union of two patterns $R_1$ and $L_2$ given their overlap. By the universal property (UP) of POs, there exists a unique arrow $m^H: H \to G_2$ that renders the diagram commutative.
This arrow gives us the PO factorization of the PB of $m_1^+$ and $m_2$. Because $m_1^+$ and $m_2$ are monos, by adhesivity, $m^H$ is also a mono (see \emph{Theorem 5.1.} in \cite{lack2005adhesive}).

Using the object $H$ we now construct two objects $P_1^H$ and $P_2^H$ given by the final PBC $P_1 \linto{p_1^H} P_1^H \lto{h_1^+} H$ to $r_1^+$ and $r^H_1$ and $P_2 \linto{p_2^H}$ $P_2^H \lto{h_2^-} H$ to $r_2^-$ and $l^H_2$ as in Diagrams \ref{composition:pbc1} and \ref{composition:pbc2}. 

\hspace{-20pt}
\begin{minipage}[t]{0.5\textwidth}
\begin{diagram}[column sep=small]\label{composition:pbc1}
	P_1 \arrow[r, "r^+_1"] \arrow[d, tail, "p_1^H"'] \& R_1 \arrow[d, tail, "r_1^H"] \\
	P_1^H \arrow[r, "h^+_1"'] \& H
\end{diagram}
\end{minipage}
\begin{minipage}[t]{0.5\textwidth}
\begin{diagram}[column sep=small]\label{composition:pbc2}
	L_2 \arrow[d, tail, "l^H_2"'] \& P_2 \arrow[l, "r_2^-"'] \arrow[d, tail, "p^H_2"] \\
	H \& P_2^H \arrow[l, "h^-_2"]
\end{diagram}\hspace{-20pt}
\end{minipage}

For the first PBC to be `meaningful', we need to make the assumption that the application of $r_1$ is \emph{reversible}. 
%More specifically, that $L_1 \linto{m_1} G_1 \lfrom{g^-_1} G_1^-$ from Diagram \ref{composition:d1} is the PO from $r_1^-$ and $m_1^-$ and $P_1 \linto{m_1^-} G^-_1 \lto{g^+_1} G_2$ is the final PBC to $r^+_1$ and $m^+_1$. 
Having made this assumption, the object $P_1^H$ can be interpreted as the result of reverting the rewrite specified by $r_1^+$ on $H$. On the other hand, the second PBC simply applies the rewrite specified by the arrow $r^-_2$ to $H$. It is easy to demonstrate that, by the UP of final PBCs, there exist unique arrows $m_1^H: P^H_1 \to G^-_1$ and $m_2^H: P^H_2 \to G^-_2$ that render Diagrams \ref{composition:d14} and \ref{composition:d141} commutative. Moreover, $m^H_1$ and $m^H_2$ are monos. 

\vspace{-10pt}

\hspace{-20pt}
\begin{minipage}[t]{0.5\textwidth}
%\vspace{10pt}
\begin{diagram}\label{composition:d14}
	P_1 \arrow[drr, tail, bend left=15, pos=0.2, "r^+_1"] \arrow[dr, "Id_{P_1}", pos=0.7] \arrow[d, "p_1^H"']  \\
	P^H_1  \arrow[drr, bend left=15, pos=0.8, "m^H \circ h_1^+" description] \arrow[dr, dashed, "m^H_1"']   \& P_1 \arrow[r, "r^+_1"] \& R_1 \arrow[d, tail, "m_1^+"] \\
	\& G_1^- \arrow[from=u, crossing over, tail, "m^-_1"'] \arrow[r, "g^+_1"'] \& G_2   \\
\end{diagram}
\end{minipage}
\begin{minipage}[t]{0.5\textwidth}
\begin{diagram}\label{composition:d141}
		   \& \& P_2 \arrow[dll, tail, bend right=15, pos=0.2, "r^-_2"'] \arrow[dl, "Id_{P_2}"', pos=0.7] \arrow[d, "p_2^H"] \\
	L_2 \arrow[d, tail, "m_2"'] \& P_2 \arrow[l, "r^-_2"'] \& P^H_2 \arrow[dll, bend right=15, pos=0.8, "m^H \circ h_2^-" description] \arrow[dl, dashed, "m^H_2"] \\
	G_2 \& G_2^- \arrow[from=u, crossing over, tail, "m^-_2"]  \arrow[l, "g^-_2"] \\
\end{diagram}
\end{minipage}

\vspace{-15pt}

To understand how the non-reversibility of the first rule prevents us from finding a `meaningful' $P_1^H$ consider the following example.

\begin{example} In Figure \ref{fig:irreversiblerule} below, we give an example expansive phase of the first rewrite applied to $G_1^-$, as in the right-most square in Diagram \ref{composition:d1} (different shapes are used to encode node identities and maps.) It selects the circle and the triangle node from $G_1^-$ and merges them. This rewrite is not reversible, i.e. we cannot restore $G_1^-$ by applying $r^+_1$ through $m_1^+$ because the depicted square does not form a PBC: the PBC of $r_1^+$ and $m_1^+$ would need an additional edge, from the circle node to the square node, in $G_1^-$.

Given the left-hand side $L_2$ of a second rule and a matching into $G_2$, we obtain the union $H$ of $R_1$ and $L_2$, given their overlap, as in Figure \ref{fig:overlap}. Reverting the rewrite specified by $r^+_1$ on $H$ gives us the object $P_1^H$ depicted in Figure \ref{fig:failcomposition}, which splits the merged circle and triangle. This splitting reconnects the black square to both circle and triangle which prevents us from constructing a match $P_1^H \into G_1^-$ necessary to obtain the desired composition---precisely because $G_1^-$ is not the PBC of $r_1^+$ and $m_1^+$.

% \vspace{-20pt}

\begin{figure}[h]
	\centering
	\begin{subfigure}[t]{0.25\textwidth}	
		\centering
		\includegraphics[scale=0.7]{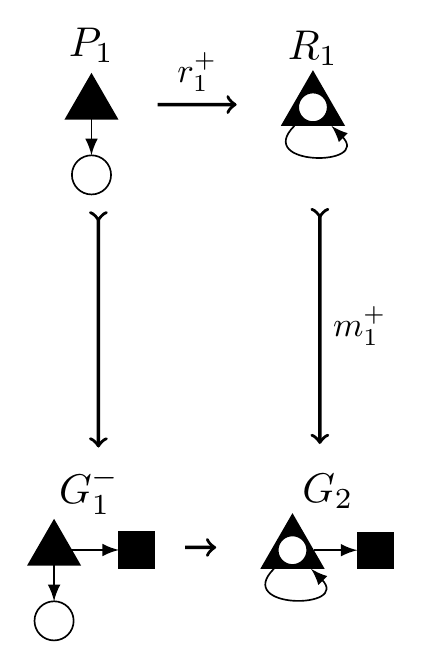}
		\caption{}\label{fig:irreversiblerule}
	\end{subfigure}
	\begin{subfigure}[t]{0.35\textwidth}
		\centering
		\includegraphics[scale=0.7]{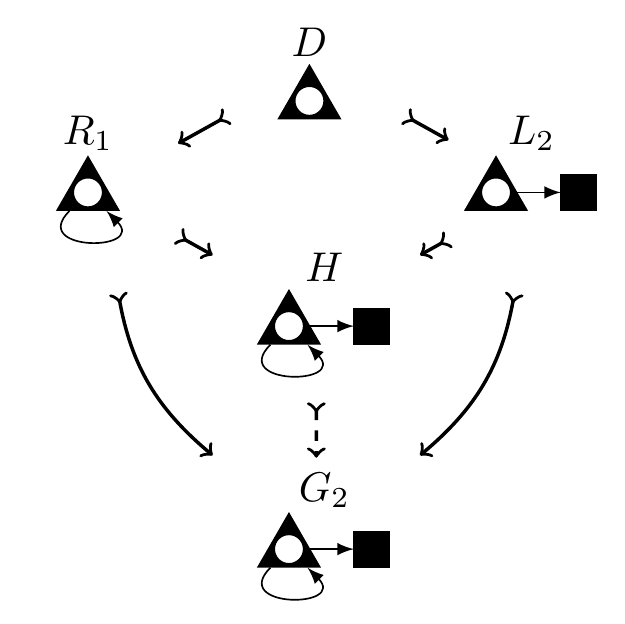}
		\caption{}\label{fig:overlap}
	\end{subfigure}
	\begin{subfigure}[t]{0.35\textwidth}
		\centering	
		\includegraphics[scale=0.7]{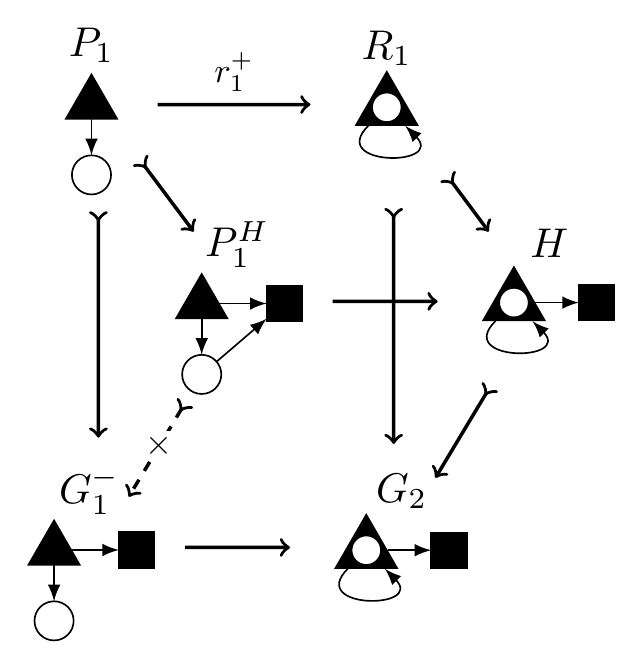}
		\caption{}\label{fig:failcomposition}
	\end{subfigure}
	\caption{Composition with an irreversible rule.}\label{fig:hierarchysideeffects}
\end{figure}

\end{example}

Next, let us construct two POs: $L_1 \linto{l_1^H} $ $L \lfrom{h^-_1} P^H_1$ from $r^-_1$ and $p_1^H$ and $R_2 \linto{r_2^H}$ $ R \lfrom{h^+_2} P^H_2$ from $r^+_2$ and $p_2^H$. The first PO reverts the rewrite of $P^H_1$ specified by $r^-_1$, and the second performs the rewrite of $P^H_2$ using $r^+_2$ and $p^H_2$. The constructed object $L$ represents the result of reversing the transformation of the pattern $P_1^H$ specified by $r_1^-$ and the instance $p^H_1$, precisely because the application of $r_1$ is reversible. By the UP of these POs we can construct unique matches $m: L \into G_1$ and $m^+: R \into G_3$.

Finally, to construct the rule composition, we find the PB $P_1^H \lfrom{p'} P \lto{p''} P_2^H$ from $h_1^+$ and $h_2^-$. The resulting rule corresponds to the span $r: L \lfrom{h_1^- \circ p'} P \lto{h_2^+\circ p''} R$. We will refer to it as the \emph{composed rule} given the overlap $o$ and write $r = \otimes (r_1, o, r_2)$.

\begin{theorem}[Synthesis]\label{synthesis}
%In a category where pushouts are stable under pullbacks (see Definition \ref{stable_pushouts})
In adhesive categories, if rewriting given by $r_1$ is reversible, application of the rule $r$ given by $\otimes (r_1, o, r_2)$ with the instance $m: L \into G_1$ produces the object $G_3$, i.e. Diagram \ref{d:synthesis} with $r^- =h_1^- \circ p'$ and $r^+ = h^+_2 \circ p''$ is an SqPO diagram. 
\end{theorem}

\begin{proposition}\label{reversiblecomposition}
In adhesive categories, the composition of two reversible rewrites is reversible.
\end{proposition}

\subsection{Hierarchies and SqPO Rewriting in Hierarchies}

A \emph{hierarchy} of objects in a category $\mathbf{C}$ is a directed acyclic graph (DAG) whose nodes are objects and whose edges are arrows from $\mathbf{C}$ such that all paths between each pair of objects are equal
\cite{harmer2020knowledge}. 
% {\color{red} if viewed as a large commutative diagrams -- homomorphisms constructed from the composition are equal.} 
We refer to the latter condition as the \emph{commutativity} condition. In the rest of this paper we assume that we are working in a fixed category $\mathbf{C}$ that has an appropriate structure for SqPO rewriting. For the commutativity of a hierarchy to be maintained, an SqPO rewrite of an object situated inside the hierarchy may require updates to other objects and arrows called \emph{propagation}.

The main model of interest to us operates on hierarchies of (simple) graphs and uses both hierarchy objects and arrows to represent knowledge. Hierarchies provide a powerful formalism for representation and update of fragmented knowledge on different interrelated abstraction levels. In this model, an edge of a hierarchy associated to an arrow $h: G \to T$ is often interpreted as \emph{typing}, i.e. the graph $T$ defines the kinds of nodes and edges that can exist in $G$. We can further interpret moving in a hierarchy along the direction of its edges as moving away from more concrete to more abstract representation. In this context, the commutativity condition guarantees that the representation of knowledge on different abstraction levels is consistent, i.e. the representation of knowledge from some concrete graph obtained by moving along alternative paths leading to the same abstract graph is consistent. The propagation framework presented in \cite{harmer2020knowledge} allows us to transform individual graphs inside a hierarchy and perform the \emph{co-evolution} of its different layers, which guarantees the consistency of knowledge at all times. 

When the knowledge represented in a hierarchy may be frequently updated by potentially different curators, it is often desirable to maintain the history of updates and be able to store multiple versions of knowledge at the same time. The design of a mathematical system providing such features, thus, constitutes the principal motivation for this work, and this system relies on the existence of a compact representation of object transformations. SqPO rewriting enables an efficient representation of object transformations using rules. To be able to build and study such a representation for transformations in hierarchies, let us briefly formulate the notions of rewriting and propagation in hierarchies (formal details can be found in \cite{harmer2020knowledge}).

Rewriting an individual object situated at a node of a hierarchy affects the objects associated at its ancestor and the descendant nodes, while the rest of the objects stay unchanged.
The restrictive phase is propagated \emph{backwards} to all the objects typed by the target of rewriting. For instance, let $G$ be an object corresponding to an ancestor of an object $T$ in a hierarchy with an associated arrow $h: G \to T$. A restrictive rewrite performing deletion and cloning of elements in $T$ induces propagation to instances of these elements in $G$. The expansive phase is propagated \emph{forward} to all the objects typing the target.  For instance, for $h: G \to T$ as before, an expansive rewrite performing addition and merging of elements in $G$ induces propagation to $T$ affecting the types of these elements. Interestingly, merging of elements in $T$ induced by such forward propagation affects all the hierarchy objects typed by $T$ (corresponding to the hierarchy side effects described in Subsection \ref{reversibleRewrSection}). The commutativity of the updated hierarchy is guaranteed by the \emph{composability conditions} imposed on the performed propagations.

\section{Rule Hierarchies}\label{s:rulehierarchy}

In this section we formulate the notion of a \emph{rule hierarchy} that serves us as a compact representation of coupled transformations of objects in a hierarchy. Intuitively, a rule hierarchy is a description of how every object situated in the hierarchy should be transformed and how the arrows between these objects should be reconstructed. In the practical application of interest, we would like to use such descriptions for recording the history of hierarchy updates. 

We then define formally how a rule hierarchy can be applied to the corresponding hierarchy of objects through specified instances and study the side-effects introduced by the application of a rule hierarchy. Such side-effects, apart from the side-effects introduced by SqPO rewriting on objects, may include some implicit changes to the arrows in the hierarchy. Thus, we formulate the conditions under which a given application of a rule hierarchy is reversible. Finally, we present how consecutive rewrites of a hierarchy can be composed. In this section we consider SqPO rewriting rules operating on objects from $\mathbf{C}$. Such rules are spans formed by objects and arrows from $\mathbf{C}$.

\begin{definition1}
A \emph{rule homomorphism} $f$ from $r_1: L_1 \lfrom{r_1^-}$ $P_1 \lto{r_1^+} R_1$ to $r_2: L_2 \lfrom{r_2^-} $ $P_2 \lto{r_2^+} R_2$ is given by three arrows $\lambda: L_1 \to L_2$, $\pi: P_1 \to P_2$ and $\rho: R_1 \to R_2$ from $\mathbf{C}$ such that $\lambda \circ r^-_1 = r_2^- \circ \pi$ and $r^+_2 \circ \pi = \rho \circ r_1^+$. 

\end{definition1}

Using rules as objects and rule homomorphisms as arrows, we obtain the category of rules $\mathbf{Rule[C]}$ over the category $\mathbf{C}$. 

\begin{definition1}
A \emph{rule hierarchy} is a hierarchy of objects in the category of rules.
\end{definition1}

Let $\mathcal{H}$ be a hierarchy of objects in $\mathbf{C}$ and $\mathcal{R}$ be a hierarchy of rules operating on objects in $\mathbf{C}$ both defined over the same DAG $\mathcal{G} = (V, E \subseteq V \times V)$. We refer to such $\mathcal{G}$ as the \emph{skeleton} of $\mathcal{H}$ and $\mathcal{R}$. For the sake of simplicity, in the rest of this section we will assume that we are working on a fixed pair $(\mathcal{H}, \mathcal{R})$ defined over the same skeleton. As a short-hand, for every node $v \in V$ we will denote the object associated to $v$ in $\mathcal{H}$ with $G_v$ and the rule associated to $v$ in $\mathcal{R}$ with $r_v: L_v \lfrom{r_v^-} P_v \lto{r_v^+} R_v$. For every edge $(s, t) \in E$ we will denote the associated arrows in $\mathcal{H}$ as $h_{(s, t)}$ and the arrows constituting the rule homomorphism in $\mathcal{R}$ as $\lambda_{(s,t)}$, $\pi_{(s,t)}$ and $\rho_{(s, t)}$.

\begin{definition1}
	An \emph{instance} of $\mathcal{R}$ in $\mathcal{H}$ is given by a function $\mathcal{I}: V \to \mathit{Monos}(\mathbf{C})$ that associates every node of the skeleton to an instance of the corresponding rule from $\mathcal{R}$ in the corresponding object from $\mathcal{H}$, i.e. $\mathcal{I}(v): L_v \into G_v$ for all $v \in V$. For every node $v \in V$ we will denote the instance $\mathcal{I}(v)$ as $m_v$.
\end{definition1}

\begin{definition1} 
$\mathcal{R}$ is \emph{applicable} to $\mathcal{H}$ through an instance $\mathcal{I}$, if for any pair of nodes $s, t \in V$ such that $(s, t) \in E$: 

\vspace{-10pt}

\hspace{-15pt}\begin{minipage}{0.55\linewidth}
\vspace{-10pt}
\begin{itemize}
	
	\item $h_{(s, t)} \circ m_s = m_t \circ \lambda_{(s, t)}$, i.e. their instances commute;

	\item if $G_s^-$ and $G_t^-$ are the results of the restrictive phase of rewriting given by the final PBC of $r^-_s$ and $m_s$, and the final PBC of $r^-_t$ and $m_t$ respectively, then there exists a unique $h_{(s, t)}^-: G_s^- \to G_t^-$ that renders Diagram \ref{applicability} commutative.

\end{itemize}
\end{minipage}
\begin{minipage}{0.35\linewidth} 
\vspace{-5pt}
\begin{equation}\label{applicability}
\hspace{5pt}
\raisebox{-30pt}{\begin{tikzcd}[ampersand replacement=\&]
		L_s  \arrow[d, tail, "m_s"'] \& P_s \arrow[l, "r_s^-"', pos=0.2] \arrow[d, tail, "m_s^-"']\arrow[ddr, "\pi_{(s, t)}" description, pos=0.2]  \\
		G_s \arrow[ddr, "h_{(s,t)}" description, pos=0.2] \& G_s^- \arrow[l, "s^-"', pos=0.2] \arrow[ddr, dashed, "h^-_{(s,t)}" description, pos=0.2] \\
		\&  L_t \arrow[from=uul, crossing over, "\lambda_{(s,t)}" description, pos=0.2] \arrow[d, tail, "m_t"] \&  P_t \arrow[l, "r_t^-"', pos=0.2, crossing over] \arrow[d, tail, "m_t^-"] \\
		\& G_t \& G_t^- \arrow[l, "t^-"'] \\ 
	\end{tikzcd}}\hspace{-50pt}
\end{equation}

\end{minipage}

\end{definition1}

\vspace{-20pt}

\begin{remark}
Observe that, if the left face in Diagram \ref{applicability} is a PB, $\mathcal{R}$ is always applicable to $\mathcal{H}$ as the unique arrow $h^-_{(s,t)}$ exists by the UP of final PBCs.
\end{remark}

To rewrite $\mathcal{H}$ using the rule hierarchy $\mathcal{R}$, applicable through an instance $\mathcal{I}$, for every node of the skeleton we simply apply the associated rule to the associated object through the instance specified by $\mathcal{I}$. To restore the arrows of $\mathcal{H}$, for every edge $(s,t) \in E$, we use the applicability condition and the UP of POs as follows. Let the back and the front faces of the cube in Diagram \ref{hierarchyapplication} be two SqPO diagrams corresponding to the above-mentioned rewriting of the objects $G_s$ and $G_t$ respectively. First of all, by the applicability of $\mathcal{R}$ given $\mathcal{I}$, there exists a unique arrow $h^-_{(s,t)}$ such that $h_{(s,t)} \circ s^- = t^- \circ h^-_{(s, t)}$ and $h^-_{(s,t)} \circ m_s^- = m_t^- \circ \pi_{(s,t)}$. This enables us to use the UP of the PO $G_s^+$ and show that there exists a unique arrow $h^+_{(s, t)}$ that renders Diagram \ref{hierarchyapplication} commutative. 

\vspace{-10pt}

\begin{equation}\label{hierarchyapplication}
    \hspace{0pt}
    \begin{tikzcd}[ampersand replacement=\&, column sep=large]
	L_s  \arrow[d, tail, "m_s"'] \& P_s \arrow[l, "r_s^-"'] \arrow[d, tail, "m^-_s"']  \arrow[r, "r_s^+"] \&  R_s \arrow[ddr,  "\rho_{(s,t)}" description, pos=0.2]  \arrow[d, tail, "m^+_s"'] \\
	G_s \arrow[ddr, "h_{(s,t)}" description, pos=0.2] \& G_s^- \arrow[l, "s^-", pos=0.8] \arrow[ddr, "h_{(s,t)}^-" description, dashed, pos=0.2] \arrow[r, "s^+", pos=0.8]  \& G_s^+  \arrow[ddr, "h_{(s,t)}^+" description, dashed, pos=0.2] \\
	\& L_t  \arrow[from=uul, crossing over, "\lambda_{(s,t)}" description, pos=0.2] \arrow[d, tail, "m_t"] \&  P_t  \arrow[from=uul, crossing over, "\pi_{(s,t)}" description, pos=0.2]  \arrow[l, crossing over, pos=0.2,  "r_t^-"'] \arrow[d, tail, "m^-_t"] \arrow[r, crossing over, pos=0.2,  "r_t^+"'] \& R_t \arrow[d, tail, "m^+_t"] \\
	\& G_t \& G_t^- \arrow[l, "t^-"'] \arrow[r, "t^+"]  \& G_t^+ \\
\end{tikzcd}
\end{equation}

\vspace{-20pt}

Therefore, the notion of a rule hierarchy can be used as a compact representation of coupled updates in hierarchies of objects. Rules describe tranformations of hierarchy objects and rule homomorphisms allow us to restore arrows between them. The commutativity condition imposed on the rule homomorphisms in a hierarchy guarantees that their application results in a valid hierarchy of objects.

\subsection{Expressing Rewriting and Propagation in Hierarchies}

In this subsection we will briefly discuss how the transformations of objects and arrows in a hierarchy $\mathcal{H}$ induced by a rewrite of an object $G$ with a rule $r: L \lfrom{r^-} P \lto{r^+} R$ through $m: L \into G$ can be represented as a rule hierarchy and its instance. Recall that, upon rewriting of an object in a hierarchy, the objects associated to ancestors and descendants of the origin of rewriting are updated according to the framework of backward and forward propagation \cite{harmer2020knowledge}, while the rest of the objects stay unaffected. We would like to construct a rule hierarchy that is defined over the skeleton of $\mathcal{H}$ and, therefore, contains rules for both affected and unaffected objects. For the sake of conciseness, here we will focus only on non-trivial updates to objects, i.e. on the construction of a rule \emph{sub-hierarchy} corresponding to the objects updated as the result of backward or forward propagation.
In this paper, we give only the high-level idea behind these constructions; full technical details can be found in \cite{oshurko2020}.

\begin{wrapfigure}{r}{0.46\textwidth}
\vspace{-25pt}
\begin{equation}\label{d:rule_lifting_full}
%\hspace{-5pt}
\begin{tikzcd}[ampersand replacement=\&]
	L_H  \arrow[d, tail, "\hat{m}"'] \& P_H \arrow[l, "\raisebox{2pt}{$\hat{r}^-$}"'] \arrow[d, tail, "\hat{m}^-"'] \arrow[r, "Id_{P_H}"] \& P_H \arrow[ddr, pos=0.6, "r^+ 
	\circ \hat{h}^-", pos=0.2]  \\
	H  \arrow[ddr, "h"', pos=0.2] \& H^- \arrow[l, pos=0.7] \arrow[ddr, "h^-", pos=0.2] \& \\
	\&  L \arrow[from=uul, "\hat{h}", crossing over, pos=0.2] \arrow[d, tail, "m"] \& P \arrow[l, "r^-", pos=0.8, crossing over] \arrow[d, tail, "m^-"] \arrow[r,  "r^+"', pos=0.8, crossing over] \arrow[from=uul,  pos=0.6, "\hat{h}^-", pos=0.2]  \&  R \arrow[d, tail, "m^+"] \\
	\& G \& G^- \arrow[l, "g^-"'] \arrow[r, "g^+"] \& G^+
\end{tikzcd}\hspace{-22pt}
\end{equation}
\vspace{-50pt}
\end{wrapfigure}
\paragraph{Backward propagation rules.} Let $H$ be an object corresponding to an ancestor of an object $G$ in a hierarchy with an associated arrow $h: H \to G$. Backward propagation of $r^-$ to $H$ can be expressed as a rule $\hat{r}: L_H \lfrom{\hat{r}^-} P_H \lto{Id_{P_H}} P_H$ with an instance $\hat{m}: L_H \into H$ that, when applied to $H$, results in an object $H^-$ that is homomorphic to the result of the original rewriting of $G$, in a way that makes Diagram 12 commute.

Such a rule, called the \emph{lifting} of $r^-$, is constructed given a specification for backward propagation (i.e. a rule factorization and a clean-up arrow \cite{harmer2020knowledge}), and contains only restrictive updates (given by deletion and cloning).
Informally, this specification indicates which changes to $G$ should be propagated to $H$ and how the updated $H$ can be `retyped' by the updated $G$. This typing is obtained as a composition $g^+ \circ h^-$ in the diagram. Together with the rule lifting, we obtain the arrows $\hat{h}: L_H \to L$ and $\hat{h}^-: P_H \to P$ defining the rule homomorphism given by $(\hat{h}, \hat{h}^-, r^+ \circ \hat{h}^-)$ as in the diagram. Therefore, the backward propagation framework allows us to extract a rule representing the specified propagation for every ancestor of the rewritten object together with homomorphisms to the original rule $r$. Moreover, under appropriate conditions (see \emph{backward composability} in \cite{harmer2020knowledge}), we can construct the homomorphisms between backward propagation rules required by the skeleton of the hierarchy.

\begin{wrapfigure}{r}{0.46\textwidth}
\vspace{-25pt}
\begin{equation}\label{forward_prop}
%\hspace{-5pt}
\begin{tikzcd}[ampersand replacement=\&]
	\&  L \arrow[d, tail, "m"] \arrow[ddl, "\hat{h}"', pos=0.2]  \& \arrow[l, "r^-"']  P \arrow[d, tail, "m^-"] \arrow[r, "r^+"] \& R  \arrow[d, tail, "m^+"] \\
	\& G \& G^- \arrow[l, "g^-", pos=0.1] \arrow[ddl, "h \circ g^-", pos=0.2] \arrow[r, "g^+", pos=0.15] \& G^+ \arrow[ddl, "h^+", pos=0.2] \\
	P_T \&   \arrow[from=uur, crossing over, "\hat{h}^-"', pos=0.2] \arrow[l, crossing over, "\raisebox{-4pt}{$Id_{P_T}$}"', pos=0.3] P_T \arrow[d, tail, "\hat{m}^-"'] \arrow[r, crossing over, "\raisebox{-4pt}{$\hat{r}^+$}", pos=0.3] \&  R_T \arrow[d, tail, "\hat{m}^+"'] \arrow[from=uur, crossing over, "\hat{h}^+"', pos=0.2] \\
	\& T \arrow[r, "t^+"'] \& T^+
\end{tikzcd}\hspace{-25pt}
\end{equation}
\vspace{-20pt}
\end{wrapfigure}
\paragraph{Forward propagation rules.} Let $T$ be an object corresponding to a descendant of an object $G$ in a given hierarchy with an associated arrow $h: G \to T$. Forward propagation of $r^+$ to $T$ can be expressed as a rule $\hat{r}: P_T \from P_T \to R_T$ with an instance $\hat{m}^-: P_T \into T$ that, when applied to $T$, results in an object $T^+$ to which the result of the original rewriting of $G$ is homomorphic, in a way renders Diagram 13 commutative. Such a rule, called the \emph{projection} of $r^+$, is constructed given a specification for forward propagation (i.e. a rule factorization and a clean-up arrow \cite{harmer2020knowledge}), and contains only expansive updates (addition and merging).

Informally, this specifies which changes to $G$ should be propagated to $T$ and how the updated $G$ can be `retyped' by the updated $T$. Together with the rule projection, we obtain three arrows $\hat{h}: L \to P_T$, $\hat{h}^-: P \to P_T$ and $\hat{h}^+: R \to R_T$ that define the rule homomorphism given by $(\hat{h}, \hat{h}^-, \hat{h}^+)$. Forward propagation thus allows us to extract a rule representing the specified propagation for every descendant of the rewritten object together with homomorphisms from the original rule $r$ and, under appropriate conditions (see \emph{forward composability} in \cite{harmer2020knowledge}), we can also construct the required rule homomorphisms.

\begin{example}
Consider the hierarchy $\mathcal{H}$ depicted in Figure \ref{ruleh}. Let $G$ be the target of rewriting with the rule highlighted in the grey area: this rule clones the circle into two semi-circles and merges one of these circles with the square node. Let the rules delimited with dashed arrows be the objects of the rule hierarchy representing respective rewriting and propagation. The rule $L_H \from P_H \to R_H$ is a backward propagation rule that specifies how the cloning in $G$ is propagated to different instances of the circle: in this case, the white circle in $H$ is to be cloned but the black circle can still be typed by the right semi-circle. The rule $L_T \from P_T \to R_T$ is a forward propagation rule that describes the merging of the nodes typing the semi-circle and the square in $G$. Finally, the hierarchy $\mathcal{H}'$ on the right represents the result of the rule hierarchy application.

\begin{figure}[h]
		\centering
		\includegraphics[scale=0.8]{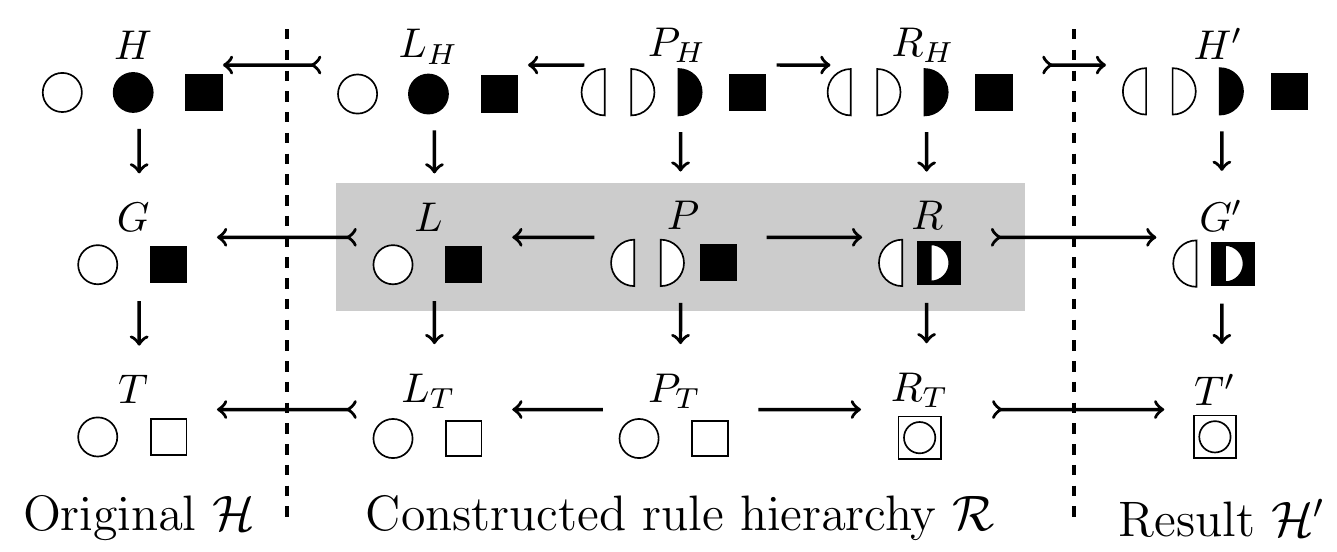}
		\caption{Example of rewriting in a hierarchy represented with a rule hierarchy.}\label{ruleh}
\end{figure}

\end{example}

\subsection{Reversible Rewriting in Hierarchies}\label{reversibleRewrSection}

In this subsection, we study the side-effects introduced by the application of a rule hierarchy. These side-effects may induce some implicit changes to the arrows representing hierarchy edges, which may prevent us from restoring the original hierarchy simply by looking at the applied rule hierarchy. In general, such side-effects make the rewriting produced by reversing the original rule hierarchy not \emph{applicable}. Let us first consider the following example.

% Due to such implicit changes, having applied a rule hierarchy, we 
\begin{example}\label{examplereversible}
	Let $G \to T$ in Figure 3a be two homomorphic objects and let $P_G \to R_G$ and $P_T \to R_T$ specify expansive phases of rules applied to these objects. The rule $P_G \to R_G$ merges one of the white circles of $G$ with the black circle; the rule $P_T \to R_T$ merges the white and black circles. As a side-effect, the other white circle is also now typed by the merged node in $T^+$: we `forget' that it was an instance of the white circle in $T$. In Figure 3b, we reverse the rules and apply them to $G^+$ and $T^+$: the merged node in $T^+$ is cloned into two circle nodes and one instance of the merged node in $G^+$ is cloned. We recover $G$ but cannot type it by $T$, precisely because we `forgot' how the circle coloured grey was typed in $T$.

	\begin{figure}[h]
		\vspace{-10pt}
		\begin{subfigure}[b]{0.5\textwidth}
		    \centering
			\includegraphics[scale=0.7]{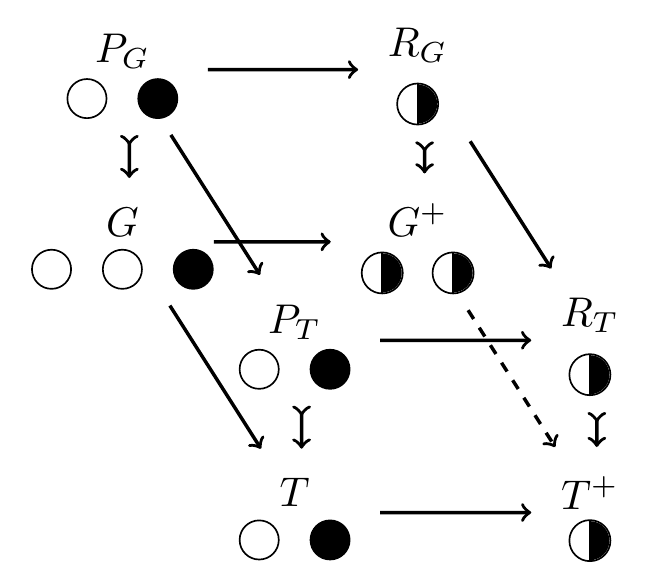}
			\caption{Application of the original rules}\label{fig:hierarchysideeffects1}
		\end{subfigure}
		\begin{subfigure}[b]{0.5\textwidth}
		    \centering
			\includegraphics[scale=0.7]{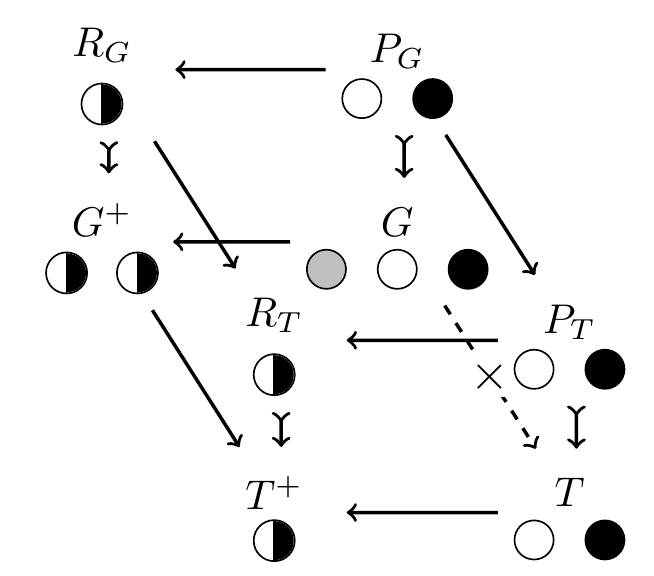}
			\caption{Application of the reversed rules}\label{fig:hierarchysideeffects2}
		\end{subfigure}
		\vspace{-5pt}
		\caption{Example of side-effects affecting hierarchy arrows}\label{fig:hierarchysideeffects}
	\end{figure}

\end{example} 

\begin{definition1}
The \emph{reverse} $\mathcal{R}^{-1}$ of $\mathcal{R}$ is the rule hierarchy whose nodes correspond to the rules $r^{-1}_v$ for all $v \in V$, and whose edges correspond to the rule homomorphisms $(\rho_{(s,t)}, \pi_{(s,t)}, \lambda_{(s,t)})$ for all edges $(s, t) \in E$.
\end{definition1}

\begin{definition1}\label{reversiblerulehierarchy} Rewriting of $\mathcal{H}$ with $\mathcal{R}$, applicable through an instance $\mathcal{I}$, is \emph{reversible}, if 	rewriting of every individual object is reversible and the reverse $\mathcal{R}^{-1}$ is applicable, i.e. for any pair of nodes $s, t \in V$ such that $(s, t) \in E$ corresponding to objects and rules as in Diagram 11, if $G_s^-$ is given as the final PBC of $r^+_s$ and $m^+_s$ and $G_t^-$ as the final PBC of $r^+_t$ and $m^+_t$, there exists a unique arrow $h_{(s,t)}^-: G_s^- \to G_t^-$ that makes the right cube in Diagram 11 commute. 
\end{definition1}

\begin{remark}
Even though the latter definition imposes rather abstract requirements, some intuitive sufficient conditions can be formulated. For example, if for every hierarchy edge the left-most face in Diagram \ref{hierarchyapplication} is a PB, the rewriting is reversible. Informally, this condition guarantees that all the instances of the elements selected by $m_t$ are also selected by $m_s$. This means that the scenario from Example \ref{examplereversible} cannot be realized, because all the instances of the merged types are selected by the rule.
\end{remark}

\subsection{Composition of Rewriting in Hierarchies}\label{s:composition}

To study composition of rewriting in a hierarchy, we will focus on a simple hierarchy with two nodes and one edge, corresponding to objects $G_1$, $T_1$ and an arrow $h_1: G_1 \to T_1$. Composition of rewriting in general hierarchies can be trivially obtained by applying this technique to every pair of hierarchy nodes connected with an edge.

Let $\mathcal{H}$ be a hierarchy corresponding to an arrow $G_1 \lto{h_1} T_1$ and let $\mathcal{R}_1$ be a rule hierarchy corresponding to rules $p_G$ and $p_T$, whose arrow $f^p: p_G \to p_T$ is given by arrows $\lambda_1$, $\pi_1$ and $\rho_1$ as in Diagram \ref{hcomposition1}. Let $G_2 \lto{h_2} T_2$ correspond to the result of the application of $\mathcal{R}_1$ through the instances $m_G$ and $m_T$ (we assume that $\mathcal{R}_1$ is applicable given $m_G$ and $m_T$). Let $\mathcal{R}_2$ be another rule hierarchy given by a homomorphic pair of rules $q_G$ and $q_T$ as in Diagram \ref{hcomposition2}. Their homomorphism $f^q: q_G \to q_T$ is given by arrows $\lambda_2$, $\pi_2$ and $\rho_2$. Let $G_3 \lto{h_3} T_3$ be the result of the application of $\mathcal{R}_2$ through the instances $n_G$ and $n_T$ as in the diagram (similarly, we assume that $\mathcal{R}_2$ is applicable given the instances).

\begin{minipage}{0.45\textwidth}
\begin{equation}\label{hcomposition1}
\hspace{-10pt}
\raisebox{-30pt}{
\begin{tikzcd}[ampersand replacement=\&]
	L_1^G \arrow[d, tail, "m_G"'] \& P_1^G \arrow[d, tail, "m_G^-"'] \arrow[l, "p_G^-"'] \arrow[r, "p_G^+"] \& R_1^G  \arrow[ddr, "\rho_1" description, pos=0.2] \arrow[d, tail, "m_G^+"'] \\ 
	G_1  \arrow[ddr, "h_1" description, pos=0.2] \& G_1^-  \arrow[ddr, "h^-_1" description, pos=0.2] \arrow[l, "g_1^-"', pos=0.2] \arrow[r, "g_1^+"', pos=0.2] \& G_2  \arrow[ddr, "h_2" description, pos=0.2] \\
	\& L_1^T \arrow[from=uul, "\lambda_1" description, pos=0.2, crossing over]  \arrow[d, tail, "m_T"]  \& P_1^T \arrow[from=uul, "\pi_1" description, pos=0.2, crossing over]   \arrow[d, tail, "m^-_T"]  \arrow[l, "p_T^-"', pos=0.2, crossing over] \arrow[r, "p_T^+"', pos=0.2, crossing over] \& R_1^T \arrow[d, tail, "m^+_T"] \\
	\& T_1 \& T_1^-  \arrow[l, "t_1^-"] \arrow[r, "t_1^+"'] \& T_2 \\
\end{tikzcd}\hspace*{-70pt}}
\end{equation}
\end{minipage}
\begin{minipage}{0.45\textwidth}
\begin{equation}\label{hcomposition2}
\hspace{-10pt}
\raisebox{-30pt}{
\begin{tikzcd}[ampersand replacement=\&]
	L_2^G \arrow[d, tail, "n_G"'] \& P_2^G \arrow[d, tail, "n_G^-"'] \arrow[l, "q_G^-"'] \arrow[r, "q_G^+"] \& R_2^G  \arrow[ddr, "\rho_2" description, pos=0.2] \arrow[d, tail, "n_G^+"'] \\ 
	G_2  \arrow[ddr, "h_2" description, pos=0.2] \& G_2^-  \arrow[ddr, "h^-_2" description, pos=0.2] \arrow[l, "g_2^-"', pos=0.2] \arrow[r, "g_2^+"', pos=0.2] \& G_3 \arrow[ddr, "h_3" description, pos=0.2] \\
	\& L_2^T \arrow[from=uul, "\lambda_2" description, pos=0.2, crossing over]  \arrow[d, tail, "n_T"]  \& P_2^T \arrow[from=uul, "\pi_2" description, pos=0.2, crossing over]   \arrow[d, tail, "n^-_T"]  \arrow[l, "q_T^-"', pos=0.2, crossing over] \arrow[r, "q_T^+"', pos=0.2, crossing over] \& R_2^T \arrow[d, tail, "n^+_T"] \\
	\& T_2 \& T_2^-  \arrow[l, "t_2^-"] \arrow[r, "t_2^+"'] \& T_3 \\
\end{tikzcd}}\hspace{-50pt}
\end{equation}
\end{minipage}

\begin{wrapfigure}{r}{0.4 \textwidth}
\vspace{-10pt}
\begin{equation}\label{rewriting_g}
\hspace{0pt}
\begin{tikzcd}[ampersand replacement=\&]
	L_X \arrow[d, tail, "l_X"] \& P_X \arrow[l, "r_X^-"'] \arrow[r, "r_X^+"] \arrow[d, tail, "l^-_X"]  \& R_X \arrow[d, tail, "l^+_X"] \\
	X_1 \& X_1^{\ominus} \arrow[l, "h_X^-"] \arrow[r, "h_X^+"'] \& X_3 \\
\end{tikzcd}\hspace{-15pt}
\end{equation}
\vspace{-30pt}
\end{wrapfigure}
We can compose these pairs of rewrites using the constructions presented in Subsection \ref{ss:composition}. Namely, if the rules $p_G$ and $p_T$ are reversible, we can find a pair of rules, $r_G:$ $L_G \lfrom{r_G^-} P_G \lto{r_G^+} R_G$ and $r_T: L_T \lfrom{r_T^-} P_T \lto{r_T^+} R_T$, and a pair of instances, $l_G: L_G \into G_1$ and $l_T: L_T \into T_1$, such that applying $r_G$ to $G_1$ and $r_T$ to $T_1$ through $l_G$ and $l_T$ respectively (as in Diagram 16, where $X$ stands for $G$ or $T$), we directly obtain $G_3$ and $T_3$ from Diagram \ref{hcomposition2}.

\begin{wrapfigure}{r}{0.3\textwidth}
\vspace{-25pt}
\begin{equation}
\hspace{2pt}
\begin{tikzcd}[ampersand replacement=\&, column sep=small]\label{hierarchyoverlap}
	\& D^G \arrow[d, dashed, "d"] \arrow[dl, tail, "x^G"'] \arrow[dr, tail, "y^G"] \\
	R^G_1 \arrow[d, "\rho_1"'] \& D^T \arrow[dl, tail, "x^T"'] \arrow[dr, tail, "y^T"] \&  L^G_2 \arrow[d, "\lambda_2"] \\
	R^T_1 \arrow[dr, tail, "m_T^+"'] \& \& L^T_2 \arrow[dl, tail, "n_T"] \\
	\& T_2 \\
\end{tikzcd}\hspace{-15pt}
\end{equation}
\vspace{-40pt}
\end{wrapfigure}
To be able to construct a rule homomorphism $f: r_G \to r_T$, we need to make the assumption that the rewriting specified by $\mathcal{R}_1$ given $m_G$ and $m_T$ is reversible, i.e. for $G_1^-$ being the PBC of $p_G^+$ and $m_G^+$  and $T_1^-$ being the PBC of $p_T^+$ and $m_T^+$, there always exists a unique arrow $h_1^-$ that renders the right-most cube in Diagram \ref{hcomposition1} commutative. Let $D^G$, $x^G$, $y^G$, $D^T$, $x^T$ and $y^T$ from Diagram 17 be the two overlaps of respectively $R^G_1$ with $L_2^G$ and $R^T_1$ with $L_2^T$, constructed as described in Subsection \ref{ss:composition} and denoted with $o^G$ and $o^T$. By the UP of PBs, there exists a unique arrow $d: D^G \to D^T$ 
%such that $\rho_1 \circ x^G = x^T \circ d$ and $\lambda$. 
using which we can construct a hierarchy of such overlaps defined over the same skeleton as $\mathcal{H}$, and together with arrows $x^G$, $y^G$, $x^T$ and $y^T$, this gives us the \emph{hierarchy overlap} $\mathcal{O}$. 

\begin{lemma}\label{rulehomexists}
If the rewriting of $\mathcal{H}$ given by $\mathcal{R}_1$ through $m_G$ and $m_T$ is reversible then, given the hierarchy overlap $\mathcal{O}$, there exists a unique rule homomorphism $f: r_G \to r_T$.
\end{lemma}

Therefore, we can construct the rule hierarchy $\mathcal{R}$ corresponding to $r_G \lto{f} r_T$. We will refer to it as the \emph{composed rule hierarchy} given the hierarchy overlap $\mathcal{O}$ and write $\mathcal{R} = \otimes(\mathcal{R}_1,\mathcal{O}, \mathcal{R}_2)$.

\begin{theorem}\label{synthesishierarchy} In an adhesive category, if the rewriting given by $\mathcal{R}_1$ is reversible, $\mathcal{R}$ is applicable given $l_G$ and $l_T$, and its application results in $G_3 \lto{h_3} T_3$. 
\end{theorem}

\begin{proposition}\label{reversiblecompositionhierarchy}
In an adhesive category, the composition of two reversible hierarchy rewrites is reversible.
\end{proposition}

\section{Transformation Audit Trails}\label{s:audit}

In this section we describe how reversibility and composition of rewriting can be used to construct the audit trail for transformations of individual objects and hierarchies of objects. The proposed framework is implemented as a part of the \texttt{ReGraph}\footnote{\url{https://github.com/Kappa-Dev/ReGraph}} Python library for building hierarchical knowledge representations based on simple graphs with attributes.

\subsection{Audit Trails for Object Transformations}\label{auditindividual}

\begin{wrapfigure}{r}{0.3\textwidth}
\vspace{-25pt}
\begin{equation}\label{rewritei}
\begin{tikzcd}[column sep=small, ampersand replacement=\&]
	L^i \arrow[d, tail, "m_i"] \& P^i \arrow[l, "r_i^-"'] \arrow[d, tail, "m^-_i"] \arrow[r, "r_i^+"] \& R^i \arrow[d, tail, "m^+_i"] \\
	G^{i-1} \& \bar{G}^{i-1} \arrow[l, "\bar{g}^-_i"] \arrow[r, "\bar{g}^+_i"'] \& G^i \\
\end{tikzcd}\hspace{-7pt}
\end{equation}
\vspace{-30pt}
\end{wrapfigure}
Let $G^0$ be the starting object whose history of transformations we would like to maintain and let $\langle r_i : L^i \lfrom{r_i^-} P^i \lto{r_i^+} R^i \mid i \in [1\ldots n] \rangle$ be a sequence of rules consecutively applied to $G^0$ through the instances $m_i: L^i \into G^{i-1}$, resulting in a sequence of objects $\langle G^i \mid i \in [1\ldots n] \rangle$ with $m_i^+: R^i \into G^i$ for $1 \leq i \leq n$, i.e. such that for every $1 \leq i \leq n$, Diagram \ref{rewritei} is a SqPO diagram. To be able to build a sound audit trail, we additionally require such a sequence of rewrites to be reversible.

\begin{definition1}
The \emph{audit trail} for the object $G^n$ consists of the sequence of rules $\langle r_i \mid i \in [1\ldots n] \rangle$ and the right-hand side instances $m_i^+: R^i \into G^i$ for $1 \leq i \leq n$.
\end{definition1}

\begin{example}\label{example1Audit}
Consider Figure 4 depicting an audit trail. Gray circles represent the states of a potentially large graph whose history of transformations we record. The small graphs inside gray circles represent the localized patterns affected by the transformations. The audit trail contains the SqPO rules representing transformations and their instances. Only the elements highlighted with solid lines are stored by the system at any time, i.e. the current state of the object, the rules and the instances representing the history of updates.

%\vspace{-5pt}

\begin{figure}[h]
	\centering
	\includegraphics[scale=0.9]{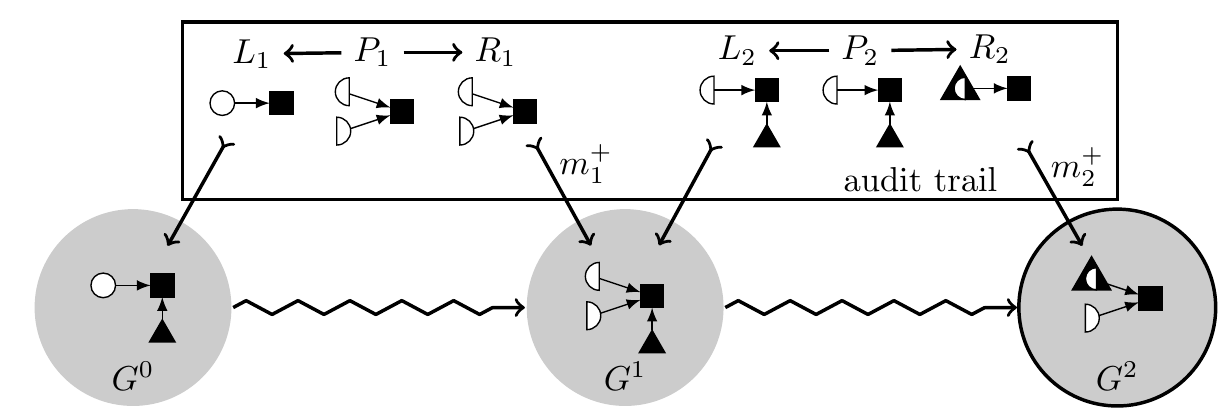}
	\caption{Example of an audit trail}\label{fig:audit}
	\vspace{-10pt}
\end{figure}

\end{example}

\begin{remark}
    The reader may wonder how rule instances (matches) are stored in the audit trail without storing the domain objects of these matches (for example, how do we store $m^+_1: R_1 \into G^1$ from Example \ref{example1Audit} without storing $G^1$ itself). In practice, we assume that graph elements (nodes and edges) are uniquely identified (e.g. with shapes and colors as in our example), and instead of storing the domain of our instances (e.g. $G^1$), it is enough to store the mapping of unique element identifiers of the co-domain and domain object constituting the rule instance (identifiers of nodes and edges of $R_1$ and $G^1$ in our example).
\end{remark}

\paragraph{Rollback.} Using such an audit trail we can \emph{rollback} to any point in the history of transformations corresponding to some intermediate object $G^i$ for $0 \leq i \leq n-1$ by applying the sequence rules $\langle r^{-1}_j \mid j \in [n \ldots i+1] \rangle$ with the corresponding instances $m_j^+: R^j \into G^j$ for $j \in [n \ldots i+1]$.

\paragraph{Maintain diverged versions.} To maintain multiple versions of an object in the audit trail, we use a technique known from VCSs as \emph{delta compression}, i.e. only the current version of the object is stored, while the other versions are encoded in a \emph{delta}, a representation of the `difference' between the versions. Let $v_1$ and $v_2$ be two versions of the starting object $G^0$ with $v_1$ being the current version. Initial delta $\Delta$ from $v_1$ to $v_2$ is set to the identity rule (the rule that does not perform any transformations) $\emptyset \lfrom{Id_{\emptyset}} \emptyset \lto{Id_{\emptyset}} \emptyset$ and the instance $u: \emptyset \into G^0$, where $\emptyset$ stands for the initial object in $\mathbf{C}$ and $u$ is the unique arrow from the initial object to $G^0$. Every rewrite of the current version of the object induces an update of the delta that consists in the composition of the previous delta and the reverse of the applied rule (recall that we assume that every rewriting in the audit trail is reversible). 

As before, let $v_1$ be the current version corresponding to some object $G$ (e.g. obtained by transforming of $G^0$) and let $r_{\Delta}: L^{\Delta} \lfrom{r_{\Delta}^-} P^{\Delta} \lto{r_{\Delta}^+} R_{\Delta}$ and $m_{\Delta}: L^{\Delta} \into G$ be respectively the rule and the instance given by $\Delta$. Let $r: L \lfrom{r^-} P \lto{r^+} R$ be a rule applied to $G$ through the instance $m: L \into G$ and $G'$ be the result of application of $r$ given $m$. To update the delta, we compute the composition $\otimes(r^{-1}, o, r^{\Delta})$ with $o$ being a span $L \lfromin{x} D \lto{y} L^{\Delta}$ obtained as a PB from $m$ and $m_{\Delta}$. The new delta is, thus, set to the rule and the instance given by the composition $\otimes(r^{-1}, o, r_{\Delta})$.

\vspace{5pt}

\begin{example}
Figure 5 illustrates how multiple version of the same object are maintained within the proposed audit trail system. Figure 1a illustrates two versions, $v_1$ and $v_2$, of the same object. As in the previous example, only the elements highlighted with solid lines are stored by the system. In this example we store the current version $v_1$ of the object and the difference between the two versions expressed with the delta. The right-most rule in Figure 1a represents a transformation of the current version $v_1$. In Figure 1b, using the overlap between the delta and the transformation rule, we compute the new delta representing the difference between the updated $v_1$ and $v_2$.

%\vspace{-10pt}

\begin{figure}[h]
	\hspace{-15pt}
	\begin{subfigure}[b]{0.45\textwidth}
	    \centering
	    \includegraphics[scale=0.9]{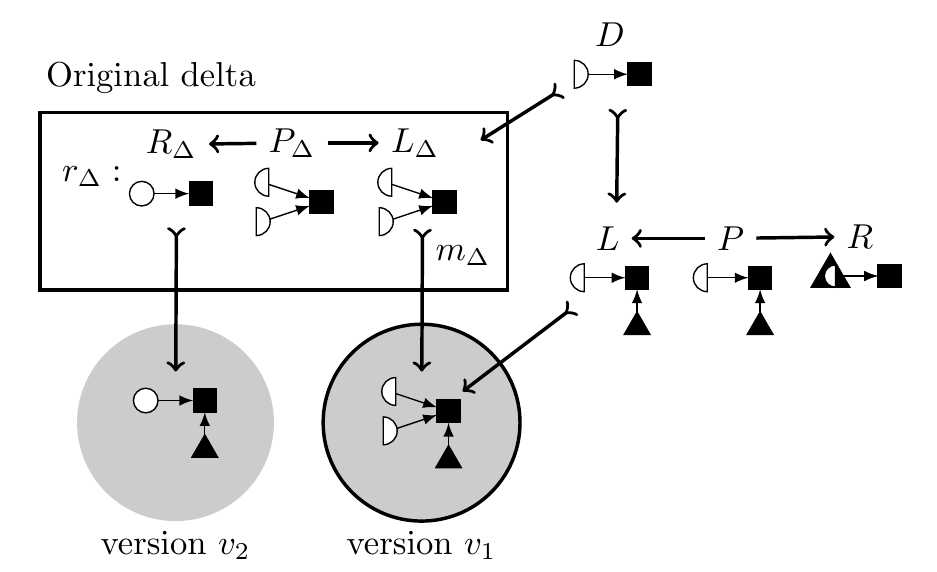}
	    \caption{}
	\end{subfigure}
	\hspace{40pt}
	\begin{subfigure}[b]{0.45\textwidth}
	    \centering
	    \includegraphics[scale=0.9]{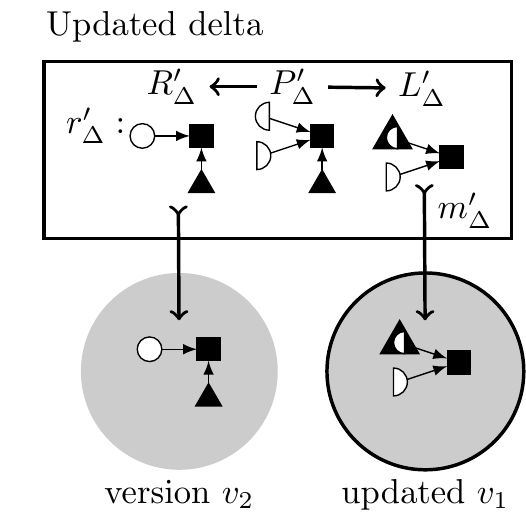}
	    \caption{}
    \end{subfigure}
    \hspace{-10pt}
	\caption{Update of a delta representing different versions of an object.}\label{deltaupd}
\end{figure}
\end{example}

%\vspace{-20pt}

\paragraph{Switch versions.} Switching between different versions of the object can be done by simply applying the rule through the instance given by the delta. Namely, if $v_1$ is the current version corresponding to an object $G$ with the delta to $v_2$ given by $\Delta = (r_{\Delta}, m_{\Delta})$, switching to $v_2$ is performed by applying $r_{\Delta}$ to $G$ through the instance $m_{\Delta}$. If $G'$ is the result of the above-mentioned rewriting and $m_{\Delta}^+: R^{\Delta} \into G'$ is its right-hand side instance, then $v_2$ becomes the current version of the object and the new delta $\Delta$ is set to $(r_{\Delta}^{-1}, m_{\Delta}^+)$.

\begin{wrapfigure}{r}{0.35\textwidth}
\vspace{-25pt}
\begin{equation}\label{canonicalmerging}
\begin{tikzcd}[ampersand replacement=\&, column sep=large, row sep=large]
	P_{\Delta} \arrow[d, "r^-_{\Delta}"'] \arrow[r, "r^+_{\Delta}"] \& R_{\Delta} \arrow[d, "\hat{r}^-"'] \\
	L_{\Delta} \arrow[ddr, tail, "m_{\Delta}" description, pos=0.2] \arrow[r, "\hat{r}^+"', pos=0.2] \& \hat{M} \arrow[ddr, tail, "\hat{m}"  description, pos=0.2] \\
	\& \bar{G} \arrow[from=uul, crossing over, tail, "m_{\Delta}^-" description, pos=0.2] \arrow[d, "g^-"] \arrow[r, crossing over, "g^+"', pos=0.2]  \& G'  \arrow[from=uul, crossing over, tail, "m_{\Delta}^+" description, pos=0.2] \arrow[d, "\hat{g}^-", dashed] \\
	\& G \arrow[r, "\hat{g}^+"] \& \hat{G}
\end{tikzcd}\hspace{-23pt}
\end{equation}
\vspace{-20pt}
\end{wrapfigure}
\paragraph{Merge versions.} Let $v_1$ be the current version corresponding to an object $G$, $v_2$ be another version corresponding to an object $G'$ and the delta between $v_1$ and $v_2$ be given by $\Delta = (r_{\Delta}, m_{\Delta})$. The left and top faces of the cube in Diagram \ref{canonicalmerging} correspond to the two phases of the application of $r_{\Delta}$ through $m_{\Delta}$. 

\emph{The canonical merging rules} for $v_1$ and $v_2$ are given by two arrows $\hat{r}^+$ and $\hat{r}^-$ constructed by the pushout $L_{\Delta} \lto{\hat{r}^+} \hat{M} \lfrom{\hat{r}^-} R_{\Delta}$ from $r_{\Delta}^-$ and $r_{\Delta}^+$ (see the back face of the cube in the diagram). The merging rule for the current object $G$ is then applied by finding the pushout $G \lto{\hat{g}^+} \hat{G} \lfromin{\hat{m}} \hat{M} $ from $m_{\Delta}$ and $\hat{r}^+$ as in the bottom face of Diagram \ref{canonicalmerging}. By the universal property of pushouts there exists a unique arrow $\hat{g}^-: G' \to \hat{G}$ that renders the cube commutative. The object $\hat{G}$ is the result of the canonical merging of $G$ and $G'$. Note that, because the application of $r_{\Delta}$ is reversible, the left face is also a pushout, which implies that the right face of the cube is also a pushout. We thus obtain the merged object $\hat{G}$ by applying the merging rule $\hat{r}^-$ to $G'$ through the instance $m_{\Delta}^+$. 

\emph{Non-canonical merging rules} are given by two arrows $r_M^+: L_{\Delta} \to M$ and $r_M^-: R_{\Delta} \to M$ such that $r_M^+ \circ r^-_{\Delta} = r_M^- \circ r^+_{\Delta}$. The merged object $G_M$ is, thus, obtained by applying the rule $r_M^+$ to $G$ through the instance $m_{\Delta}$ or the rule $r_M^-$ to $G'$ through the instance $m^+_{\Delta}$.

\begin{example}

Figure 6 illustrates how the operation of merge can be performed within the audit trail. The delta depicted in the figure corresponds to the delta from Figure \ref{deltaupd}. Using this delta, we compute the canonical merging rules for the two versions given by $L_{\Delta} \to \hat{M}$ and $R_{\Delta} \to \hat{M}$ respectively. In this example, the merging rule merges the two semi-circles produced by the cloning in $v_1$ (see the first rule applied tot $G^0$ in Figure \ref{fig:audit}), which results into the object represented with gray solid circle.
\end{example}

\vspace{-15pt}

\begin{figure}[h]
	\centering
    \includegraphics[scale=0.9]{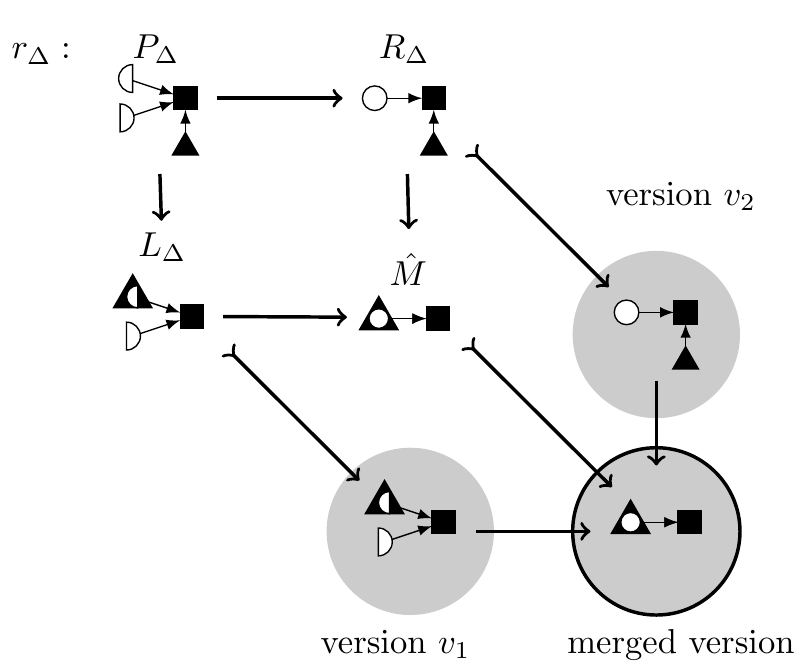}
	\caption{Merging different versions of an object.}\label{fig:mergeex}
	\vspace{-10pt}
\end{figure}

\subsection{Audit Trails for Hierarchy Transformations}

Let $\mathcal{H}^0$ be the starting hierarchy of objects, defined over a skeleton $\mathcal{G} = (V, E)$, whose history of transformations we would like to maintain. Let $\langle \mathcal{R}_i \mid i \in [1\ldots n] \rangle$ be a sequence of rule hierarchies consecutively applied to $\mathcal{H}^0$ through the instances $\mathcal{I}_i$, resulting in a sequence of hierarchies $\langle \mathcal{H}^i \mid i \in [1\ldots n] \rangle$ with the right-hand side instances given by $\mathcal{I}_i^+$  for $1 \leq i \leq n$, i.e. for every $v \in V$, $\mathcal{I}_i^+(v): R_v^i \into G^i_v$. As in the case of individual objects, to be able to build an audit trail, we require all the rewrites to be reversible. 

\begin{definition1}
The \emph{audit trail} for $\mathcal{H}^n$ consists of the sequence of rule hierarchies $\langle \mathcal{R}_i \mid i \in [1\ldots n] \rangle$ and the right-hand side instances $\mathcal{I}_i^+$ for $1 \leq i \leq n$.
\end{definition1} 

\paragraph{Rollback.} Using the audit trail we can \emph{rollback} to any point in the history of transformations corresponding to some intermediate hierarchy $\mathcal{H}^i$ for $0 \leq i \leq n-1$. This can be done by applying the rule hierarchies $\langle \mathcal{R}^{-1}_j \mid j \in [n \ldots i+1] \rangle$ with the corresponding instances $\mathcal{I}_j^+$, where $\mathcal{I}_j^+(v): R^j_v \into G^j_v$ for every $v \in V$ and $j \in [n \ldots i+1]$.

\paragraph{Maintain diverged versions.} To accommodate multiple versions of a hierarchy, we use delta compression. Let $v_1$ and $v_2$ be two versions of the starting hierarchy $\mathcal{H}^0$ with $v_1$ being the current version. The initial delta $\Delta$ from $v_1$ to $v_2$ is set to the identity rule hierarchy with the rule $\emptyset \lfrom{Id_{\emptyset}} \emptyset \lto{Id_{\emptyset}} \emptyset$ at every node $v \in V$. We set the instance $\mathcal{I}(v)$ for every $v \in V$ to be the unique arrow $u_v: \emptyset \into G_v^0$. Every rewrite of the current version of the hierarchy induces an update of the delta that consists in the composition of the previous delta and the reverse of the applied rule hierarchy. Let $v_1$ be the current version corresponding  to some hierarchy $\mathcal{H}$.

\begin{wrapfigure}{r}{0.3\textwidth}
\vspace{-15pt}
\begin{equation}
\hspace{0pt}
\begin{tikzcd}[ampersand replacement=\&, column sep=small, row sep=small]\label{hierarchyversioningoverlap}
	\& D_s \arrow[d, dashed, pos=0.8] \arrow[dl, tail, "x_s"'] \arrow[dr, tail, "y_s"] \\
	L_s \arrow[d, "\lambda_{(s,t)}"']  \& D_t \arrow[dl, tail, "x_t"] \arrow[dr, tail, "y_t"] \& L_s^{\Delta} \arrow[d, "\lambda_{(s,t)}^{\Delta}"] \\
	L_t \arrow[dr, tail, "m_t"'] \& \& L_t^{\Delta} \arrow[dl, tail, "m_t^{\Delta}"] \\
	\& G_t \\
\end{tikzcd}\hspace{-10pt}
\end{equation}
\vspace{-30pt}
\end{wrapfigure}
Let $\mathcal{R}_{\Delta}$ and $\mathcal{I}_{\Delta}$ be the rule hierarchy and instance given by $\Delta$, where $r^{\Delta}_v: L_v^{\Delta} \lfrom{r_{v, \Delta}^-} P_v^{\Delta} \lto{r_{v, \Delta}^+} R_v^{\Delta}$ and
$m_v^{\Delta}: L_v^{\Delta} \into G_v$ are the rule and the instance corresponding to a node $v \in V$. Let $\mathcal{R}$ be a rule hierarchy applied to $\mathcal{H}$ through the instance $\mathcal{I}$ and $\mathcal{H}'$ be the result of the corresponding rewriting. The new delta is given by the rule hierarchy and the instance obtained as the composition $\otimes(\mathcal{R}^{-1}, \mathcal{O}, \mathcal{R}^{\Delta})$ with $\mathcal{O}$ being the hierarchy overlap computed by finding the overlaps between $L_v$ and $L_v^{\Delta}$ for every node $v \in V$ and the arrow $D_s \to D_t$ between overlaps given by the UP of PBs as in Diagram \ref{hierarchyversioningoverlap} for every edge $(s,t) \in E$.

\paragraph{Switch versions.} Switching between versions of the hierarchy is performed by applying the rule hierarchy though the instance given by the delta. If $v_1$ is the current version corresponding to a hierarchy $\mathcal{H}$ with the delta given by $\Delta = (\mathcal{R}_{\Delta}, \mathcal{I}_{\Delta})$, switching to $v_2$ is performed by applying $\mathcal{R}_{\Delta}$ to $\mathcal{H}$ through $\mathcal{I}_{\Delta}$. For $\mathcal{H}'$ being the result of rewriting and $\mathcal{I}_{\Delta}^+$ being its right-hand side instance (where for every $v \in V$, $\mathcal{I}_{\Delta}^+(v): R_v^{\Delta} \into G_v'$), $v_2$ becomes the current version of the object and the new delta is $(\mathcal{R}_{\Delta}^{-1}, \mathcal{I}_{\Delta}^+)$.

\paragraph{Merge versions.} Let $v_1$ be the current version corresponding to a hierarchy $\mathcal{H}$, $v_2$ be another version corresponding to a hierarchy $\mathcal{H}'$ and the delta between $v_1$ and $v_2$ be given by $\Delta = (\mathcal{R}_{\Delta}, \mathcal{I}_{\Delta})$. %Merging the two diverged versions $v_1$ and $v_2$ can be performed in a canonical and non-canonical fashion.

\emph{The canonical merging rule hierarchy} can be constructed in the following way. For every individual hierarchy node we construct the canonical merging rule according to the framework described in Subsection \ref{auditindividual}. Let the back and front faces of the cube in Diagram \ref{mergehoms} correspond to the POs defining pairs of merging rules corresponding to nodes $s, t \in V$ such that $(s, t) \in E$. We can apply the universal property of the POs and show that there exists a unique arrow $m_{(s, t)}: \hat{M}_s \to \hat{M}_t$ that makes the diagram commute. The merging rule hierarchy $\hat{\mathcal{R}}^+$ for $\mathcal{H}$ is, thus, given by rules $L_v \lfrom{Id_{L_v}} L_v \lto{\hat{r}_v^+} \hat{M}_v$, for all $v \in V$, and rule homomorphisms defined by arrows $(\lambda_{(s, t)}, \lambda_{(s, t)}, m_{(s, t)})$,  for all $(s, t) \in E$. On the other hand, the merging rule hierarchy for $\mathcal{H}'$ is given by rules $R_v \lfrom{Id_{R_v}} R_v \lto{\hat{r}_v^-} \hat{M}_v$, for all $v \in V$, and rule homomorphisms defined by arrows $(\rho_{(s, t)}, \rho_{(s, t)}, m_{(s, t)})$,  for all $(s, t) \in E$. Let $\hat{G}_s$ and $\hat{G}_t$ be the result of merging corresponding to the nodes $s$ and $t$. By the universal property of POs there exists a unique arrow $\hat{h}_{(s, t)}$ that renders Diagram \ref{mergehomsgraphs} commutative. Therefore, using such objects $\hat{G}_v$ for every $v \in V$ and arrows $h_{(s,t)}$ for every $(s, t) \in E$, we can construct the hierarchy $\hat{\mathcal{H}}$ corresponding to the result of canonical merging of $\mathcal{H}$ and $\mathcal{H}'$. \emph{Non-canonical merging} can be specified using a hierarchy of objects $\mathcal{M}$ defined over the same skeleton as $\mathcal{H}$, and a pair of arrows $\bar{r}_v^+: L_v \to M_v$ and $\bar{r}_v^-: R_v \to M_v$ such that $\bar{r}_v^+ \circ r^-_v = \bar{r}_v^- \circ r^+_v$ for every $v \in V$.

\vspace{-10pt}

\hspace{-25pt}
\begin{minipage}[t]{0.5\textwidth}
\begin{equation}\label{mergehoms}
\begin{tikzcd}[ampersand replacement=\&, column sep=large]
P_s \arrow[d, "r^-_s"'] \arrow[r, "r^+_s"] \& R_s \arrow[d, "\hat{r}^-_s"'] \\
L_s \arrow[ddr, "\lambda_{(s, t)}" description, pos=0.2] \arrow[r, "\hat{r}^+_s"', pos=0.2] \& \hat{M}_s \arrow[ddr, "m_{(s, t)}" description, pos=0.2, dashed] \\
\& P_t \arrow[from=uul, "\pi_{(s, t)}" description, pos=0.2, crossing over]  \arrow[d, "r^-_t"] \arrow[r, "r^+_t"', pos=0.2] \& R_t \arrow[from=uul, "\rho_{(s, t)}" description, pos=0.2, crossing over]  \arrow[d, "\hat{r}^-_t"] \\
\& L_t \arrow[r, "\hat{r}^+_t"] \& \hat{M}_t \\
\end{tikzcd}\hspace{-23pt}
\end{equation}
\end{minipage}
\begin{minipage}[t]{0.5\textwidth}
\begin{equation}\label{mergehomsgraphs}
\begin{tikzcd}[ampersand replacement=\&, column sep=large]
L_s \arrow[d, tail, "m_s"'] \arrow[r, "\hat{r}^+_s"] \& \hat{M}_s \arrow[d, tail, "\hat{m}_s"'] \\
G_s \arrow[ddr, "h_{(s, t)}" description, pos=0.2] \arrow[r, "\hat{g}^+_s"', pos=0.2] \& \hat{G}_s \arrow[ddr, "\hat{h}_{(s, t)}" description, pos=0.2, dashed] \\
\& L_t \arrow[from=uul, "\lambda_{(s, t)}" description, pos=0.2, crossing over]  \arrow[d, tail, "m_t"] \arrow[r, "r^+_t"', pos=0.2] \& \hat{M}_t \arrow[from=uul, "m_{(s, t)}" description, pos=0.2, crossing over]  \arrow[d, tail, "\hat{m}_t"] \\
\& G_t \arrow[r, "\hat{g}^+_t"] \& \hat{G}_t \\
\end{tikzcd}\hspace{-23pt}
\end{equation}
\end{minipage}

\section{Conclusions}
\vspace{-3pt}

In this paper we have described how the reversibility and composition of SqPO rewriting can be used to design an audit trail framework for individual objects and hierarchies of objects.
In particular, we have presented the construction that composes consecutive SqPO rewrites, provided the first rewrite is reversible. We have also presented the notion of rule hierarchy, that generalizes SqPO rewriting to hierarchies of objects, and allows for an efficient representation of rewriting and its propagation, as previously presented in \cite{harmer2020knowledge}. We have studied the conditions under which an arbitrary rule hierarchy can be applied to the corresponding hierarchy of objects and described the conditions for such an application to be reversible. We then briefly discussed the construction that can be used to compose consecutive applications of two rule hierarchies. Finally, we have described how an audit trail for individual objects and hierarchies of objects can be defined. Such an audit trail allows for the maintenance of the history of transformations and provides the means for reverting sequences of such transformations. 
Moreover, it enables the accommodation of multiple versions of the same object that have diverged as the result of conflicting rewrites.

As future work, we would like to study how an arbitrary transformation from a sequence of rewrites can be undone for  individual objects and hierarchies of objects. This question is directly related to the theory of \emph{causality} for SqPO rewriting and requires a generalization for such rewriting in hierarchies.

\nocite{*}
\bibliographystyle{eptcs}
\bibliography{generic}
\end{document}